\documentclass[12pt,preprint]{aastex}
\usepackage{url}
\usepackage{multirow}
\usepackage{enumerate}
\usepackage{hyperref}

\newcommand{\sat}[1]{\it\uppercase{#1}\rm}
\newcommand{\fig}[1]{Figure~\ref{#1}}

\newcommand{\speed}[1]{#1 km~s${}^{-1}$}
\newcommand{\aspeed}[1]{$\sim\,$#1 km~s${}^{-1}$}

 \usepackage{color}           
 \definecolor{DarkGreen}{rgb}{0.0,0.45,0.0}  

\begin{document}

\shorttitle{Contraction \& Eruption} %

\shortauthors{Liu et al.}

\title{Contracting and Erupting Components of Sigmoidal Active Regions}

\author{Rui Liu\altaffilmark{1, 2}, Chang Liu\altaffilmark{2}, Tibor T\"{o}r\"{o}k\altaffilmark{3}, Yuming Wang\altaffilmark{1}, \& Haimin Wang\altaffilmark{2}}

\altaffiltext{1}{CAS Key Lab of Geospace Environment, Department of Geophysics \& Planetary Sciences, University of Science \& Technology of China, Hefei 230026, China}

\altaffiltext{2}{Space Weather Research Laboratory, Center for Solar-Terrestrial Research, NJIT, Newark, NJ
07102, USA}

\altaffiltext{3}{Predictive Science Inc., 9990 Mesa Rim Road, Suite 170, San Diego, CA 92121, USA}
\email{rliu@ustc.edu.cn}

\begin{abstract}
It is recently noted that solar eruptions can be associated with the contraction of coronal loops that are not involved in magnetic reconnection processes. In this paper, we investigate five coronal eruptions originating from four sigmoidal active regions, using high-cadence, high-resolution narrowband EUV images obtained by the Solar Dynamic Observatory (\sat{sdo}). The magnitudes of the flares associated with the eruptions range from the \sat{goes}-class B to X. Owing to the high-sensitivity and broad temperature coverage of the Atmospheric Imaging Assembly (AIA) onboard \sat{sdo}, we are able to identify both the contracting and erupting components of the eruptions: the former is observed in cold AIA channels as the contracting coronal loops overlying the elbows of the sigmoid, and the latter is preferentially observed in warm/hot AIA channels as an expanding bubble originating from the center of the sigmoid. The initiation of eruption always precedes the contraction, and in the energetically mild events (B and C flares), it also precedes the increase in \sat{goes} soft X-ray fluxes. In the more energetic events, the eruption is simultaneous with the impulsive phase of the nonthermal hard X-ray emission. These observations confirm the loop contraction as an integrated process in eruptions with partially opened arcades. The consequence of contraction is a new equilibrium with reduced magnetic energy, as the contracting loops never regain their original positions. The contracting process is a direct consequence of flare energy release, as evidenced by the strong correlation of the maximal contracting speed, and strong anti-correlation of the time delay of contraction relative to expansion, with the peak soft X-ray flux. This is also implied by the relationship between contraction and expansion, i.e., their timing and speed.

\end{abstract}


\section{Introduction}

It is generally regarded that solar eruptions are due to a disruption of the force balance
between the upward magnetic pressure force and the downward magnetic tension force. Since the
eruption can only derive its energy from the free energy stored in the coronal magnetic field
\citep{forbes00}, ``the coronal field lines must contract in such a way as to reduce the magnetic
energy $\int_\mathcal{V} B^2/8\pi$'' \citep{hudson00}. The contraction must be associated with
the reduction of the magnetic tension force for each individual loop-like field line undergoing
contraction, as its footpoints are effectively anchored in the photosphere. Eventually a new
force balance would be achieved between the magnetic pressure and tension force after the energy
release. From an alternative viewpoint, the average magnetic pressure $B^2/8\pi$ must decrease
over the relevant volume $\mathcal{V}$ across the time duration of the eruption. $\mathcal{V}$
can be roughly regarded as the flaring region, primarily in which magnetic energy is converted
into other forms of energies. The contraction process, termed as ``magnetic implosion'' by
\citet{hudson00}, is very similar to the shrinkage of post-flare loops \citep{fa96}, except that
loop shrinkage is driven by temporarily enhanced magnetic tension force at the cusp of the newly
reconnected field lines, whereas loop contraction by reduced magnetic pressure in the flaring
region. Additionally, with newly reconnected loops piling up above older ones, the post-flare
arcade as a whole often expands rather than shrinks with time.

\begin{figure}\epsscale{0.8}%
\plotone{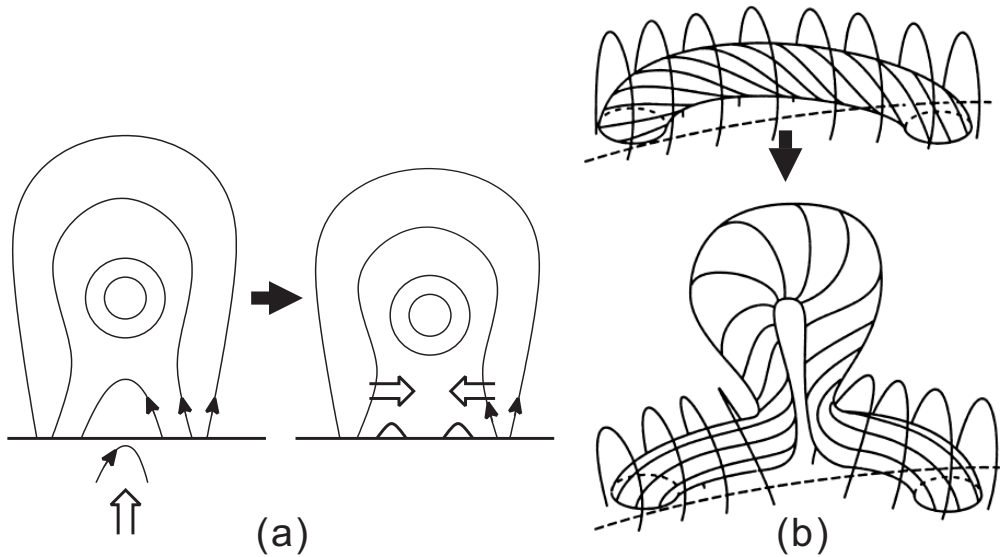}
\caption{CME models relevant to magnetic implosion. \textit{a}) Schematic diagram of the emerging
flux triggering mechanism for CMEs \citep[adapted from][]{cs00}. The emerging flux inside the
filament channel cancels the pre-existing loops, which results in the in-situ decrease of the
magnetic pressure. Magnetized plasmas are driven inward to form a current sheet beneath the flux
rope; \textit{b}) Schematic sketch showing that in the three-dimensional space a twisted flux
rope can rupture the overlying magnetic arcade and erupt by pushing the magnetic arcade aside
\citep[adapted from][]{sturrock01}. With the escape of the flux rope, the arcade field undergoes
a contraction due to the decreased magnetic pressure in the core field. \label{model}}
\end{figure}

\citet{hudson00} concluded that ``a magnetic implosion must occur simultaneously with the energy
release'' , based on no assumption about the energy release process itself. However, the detailed
timing and location of loop contraction might provide diagnostic information on the eruption mechanism.
For example, when the reconnection-favorable flux emerges inside a filament channel \citep[Figure
\ref{model}(a); adapted from][]{cs00}, it cancels the small magnetic loops below the flux rope,
which results in a decrease of the local magnetic pressure. The whole dipolar magnetic structure
must contract correspondingly. Meanwhile, plasmas on both sides of the polarity-inversion
line (PIL) would move inward to form a current sheet below the flux rope and the subsequent
evolution could follow the paradigm of the standard flare model \citep[e.g.,][]{kp76}. In that
case, overlying coronal loops could be observed to initially contract and then erupt. In a
different scenario, a twisted flux rope confined by potential-like magnetic fields is found to be
energetically favorable to ``rupture'' through the overlying arcade via ideal-MHD processes
\citep[Figure \ref{model}(b); adapted from ][]{sturrock01}. This is clearly demonstrated in MHD
simulations by \citet{gf06} and \citet{rachmeler09}, in which overlying loops can be seen to be
pushed upward and aside as the flux rope kinks and expands, and after the rope ruptures through
the arcade, overlying loops on both sides quickly contract toward the core region, due to the
reduction of the magnetic pressure in the core field with the escape of the flux rope. In
particular for this scenario (\fig{model}(b)), one would expect to see both the expanding flux rope
and the contracting overlying loops during the eruption as long as the arcade is only partially opened.
Although both scenarios involve a pre-existent flux rope, supposedly they can also accommodate
those models in which the flux rope forms immediately prior to \citep[e.g.,][]{moore01}, or
during the course of \citep[e.g.,][]{adk99}, the eruption.


Corresponding to the aforementioned models (\fig{model}), our previous observational studies also
suggest two different scenarios, i.e., 1) the bunch of coronal loops undergoing contraction later
becomes the front of the eruptive structure \citep{lwa09}; and 2) the contracting loops are
distinct from the eruptive structure \citep{lw09, lw10}. The role of contraction in the eruption, however, has been unclear in both scenarios. For Scenario 1, the event reported by \citet{lwa09} remains
unique in the literature; as for Scenario 2, the eruptive structure is not easy to detect before
its appearance as a CME in coronagraph, unless there is dense filament material serving as the
tracer \citep{lw09}. In some cases its slow ascension and expansion during the early stage might
manifest as the gradual inflation of overlying coronal loops \citep{liu10a}. Only with the
advent of the Solar Dynamic Observatory \citep[\textit{SDO};][]{pesnell12} which provides a continuous and wide
temperature coverage, is the eruptive structure itself more frequently identified beneath the
coronagraph height as a hot, diffuse plasmoid \citep[e.g.,][]{liu10b, cheng11}. 

Here in a further investigation of Scenario 2, we identify both the erupting and
contracting components using \sat{sdo} data, hence for the first time we are able to study in
detail their relationship as well as the implication for the eruption mechanism and the
associated energy release process.  In the rest of the paper, we present in Section 2 the results
of the investigation on five flares (Table 1) observed by the Atmospheric Imaging Assembly
\citep[AIA;][]{lemen12} onboard \sat{sdo}, and we make concluding statements in Section 3.

\begin{figure}\epsscale{0.85}
\plotone{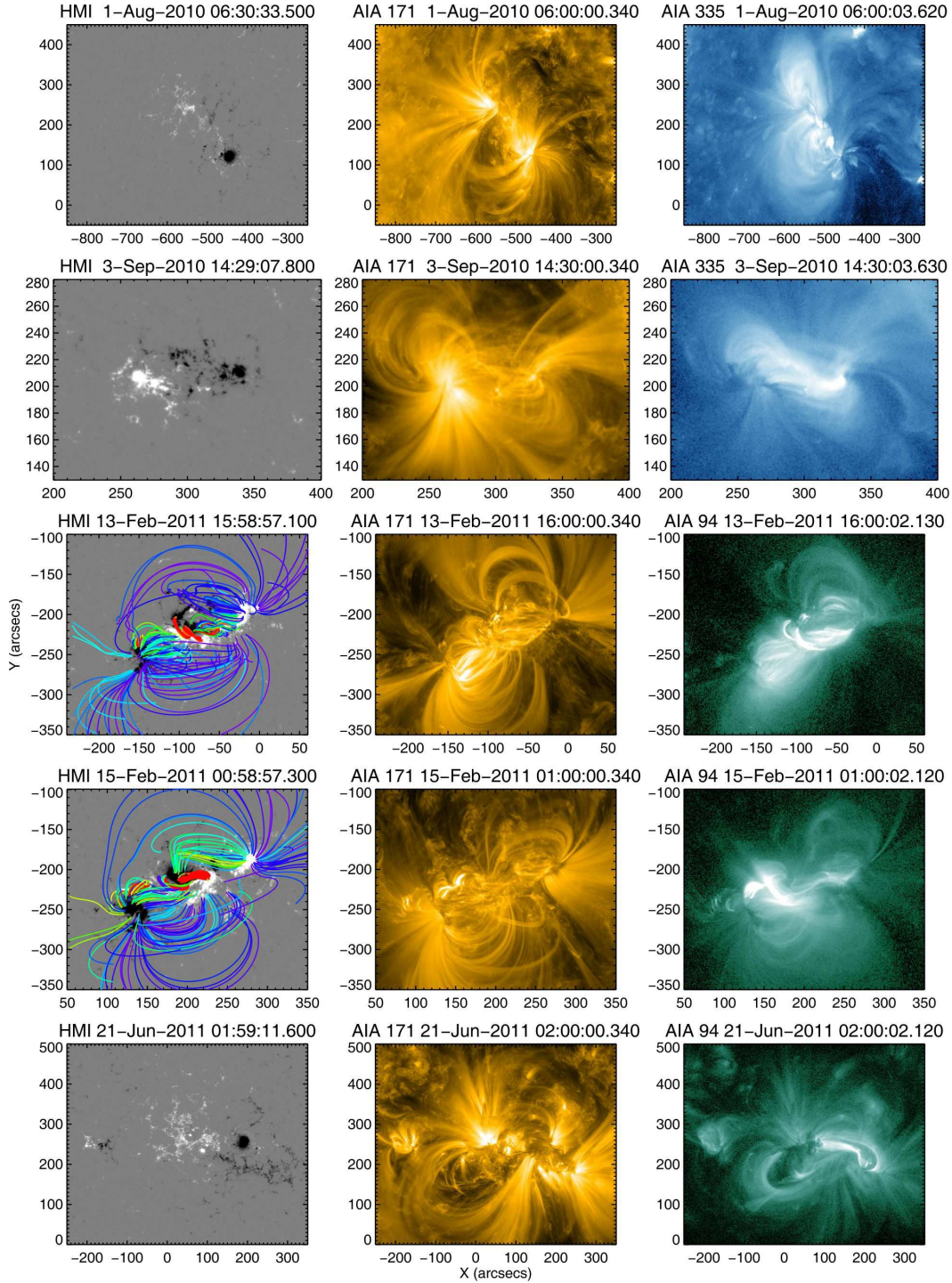}
\caption{Pre-flare configuration for the five flares studied. Left column: line-of-sight
magnetograms obtained by the Helioseismic and Magnetic Imager (HMI) onboard \sat{sdo}; Middle and
Right columns: corresponding EUV images in the cold and warm/hot AIA channels, respectively,
showing the sigmoidal morphology and structure. For AR 11158 (3rd and 4th rows), we use HMI
vector magnetograms to construct nonlinear force-free field (NLFFF; see the text for details).
The extrapolated field lines are color-coded according to the intensity of vertical currents on
the surface. \label{sigmoid}}
\end{figure}

\section{Observation}

\subsection{Overview}
\begin{deluxetable}{cccccccc}
\tablecolumns{8} %
\tabletypesize{\small} %
\tablewidth{0pt} %
\tablecaption{List of Events}%
\tablehead{\colhead{Date} & \colhead{AR} & \colhead{Location} & \colhead{Hale} &
\colhead{GOES} & \colhead{$v_c^{max}$ (km~s${}^{-1}$)\tablenotemark{a}} & \colhead{$v_e^{max}$ (km~s${}^{-1}$)\tablenotemark{b}} & \colhead{$\Delta t$ (min)\tablenotemark{c} } }%
\startdata

2010-08-01 & 11092 & N13E21 & $\alpha/\beta$ & C3.2 & -51 & 83 & 9.0 \\

2010-09-03 & 11105 & N19W23 & $\beta/-$ & B2.8 & -12 & 94 & 34.6 \\

2011-02-13 & 11158 & S19W03 & $\beta/\beta$ & M6.6 & -195 & 538 & 1.8 \\

2011-02-15 & 11158 & S21W21 & $\beta\gamma/\beta\gamma$ & X2.2 & -320 &  401 & 2.4 \\

2011-06-21 & 11236 & N17W19 & $\beta\gamma/\beta\gamma$ & C7.7 & -57 & 90 & 12.4
\enddata
\tablenotetext{a}{Maximum contracting speed}
\tablenotetext{b}{Maximum expanding speed}
\tablenotetext{c}{Time delay of contraction relative to expansion}
\end{deluxetable}

In addition to the symbiosis of the erupting and contracting component, the five
flares studied here all occurred in sigmoidal active regions (\fig{sigmoid}), which took a
sinusoidal shape in the warm AIA channels such as 211~\AA\ (dominated by \ion{Fe}{14}, $\log
T=6.3$) and 335~\AA\ (dominated by \ion{Fe}{17}, $\log T=6.4$) or hot channels like 94 \AA\
(dominated by \ion{Fe}{18}, $\log T=6.8$). By close inspection, one can see that two groups of
J-shaped loops which are oppositely oriented with respect to each other collectively make the
sinusoidal appearance (\fig{sigmoid}). In cold channels such as 171~\AA\ (dominated by
\ion{Fe}{9} and \ion{Fe}{10}, $\log T=5.8$) and 193~\AA\ (dominated by \ion{Fe}{12}, $\log
T=6.1$), these regions were dominated by large-scale loops arched over the elbows of the hot
sigmoid, suggesting that the highly sheared core field is restrained by the potential-like
overlying field. Since nonpotential (sheared or twisted) fields are reservoir for magnetic free
energy, it is not surprising that sigmoidal regions are significantly more likely to be eruptive
than non-sigmoidal regions \citep{hudson98, chm99, glover00}, and are deemed as one of the most
important precursor structures for solar eruptions.


Of the five flares, both the M6.6 flare on 2011 February 13 and the X2.2 flare two days later on
February 15 occurred in the same active region 11158. One can see that on February 13 when it was
still classified as a $\beta$-region, AR 11158 was only a ``rudimentary'' sigmoid compared with
its status on February 15. The hot loops in AIA 94~\AA\ in the center of the active region,
however, were already highly sheared, taking the similar east-west orientation as the major PIL
along which the two bipolar regions interacted and major flares took place \citep[see][for
details]{bvd12}.

Utilizing the newly released vector magnetograms with the $0''.5$ pixel size for AR 11158
\citep{hoeksema11} obtained by the Helioseismic and Magnetic Imager \citep[HMI;][]{scherrer12}
onboard \sat{sdo}, we constructed the nonlinear force-free field (NLFFF) model using the
``weighted optimization'' method \citep{wieg04} after preprocessing the photospheric boundary to
best suit the force-free condition \citep{wis06}. NLFFF extrapolation using the vector
magnetogram at about 16:00 UT on 2011 February 13 indeed shows highly sheared field lines near
the flaring PIL and potential-like field lines overlying it, similar in morphology to the hot and
cold coronal loops, respectively (\fig{sigmoid}). NLFFF result using the vector magnetogram at
about 01:00 UT on February 15 gives a similar result. The extrapolated field lines are
color-coded according to the intensity of vertical currents on the surface. Field lines whose
footpoints are associated with strong current densities ($>0.02$ A m${}^{-2}$) are in red colors.
The footpoints of these red field lines are co-spatial with the four footpoint-like flare
brightenings in AIA 94 \AA\ images \citep{liuc12, wangs12}.

\subsection{2010 August 1 Event}
\begin{figure}\epsscale{1}
\plotone{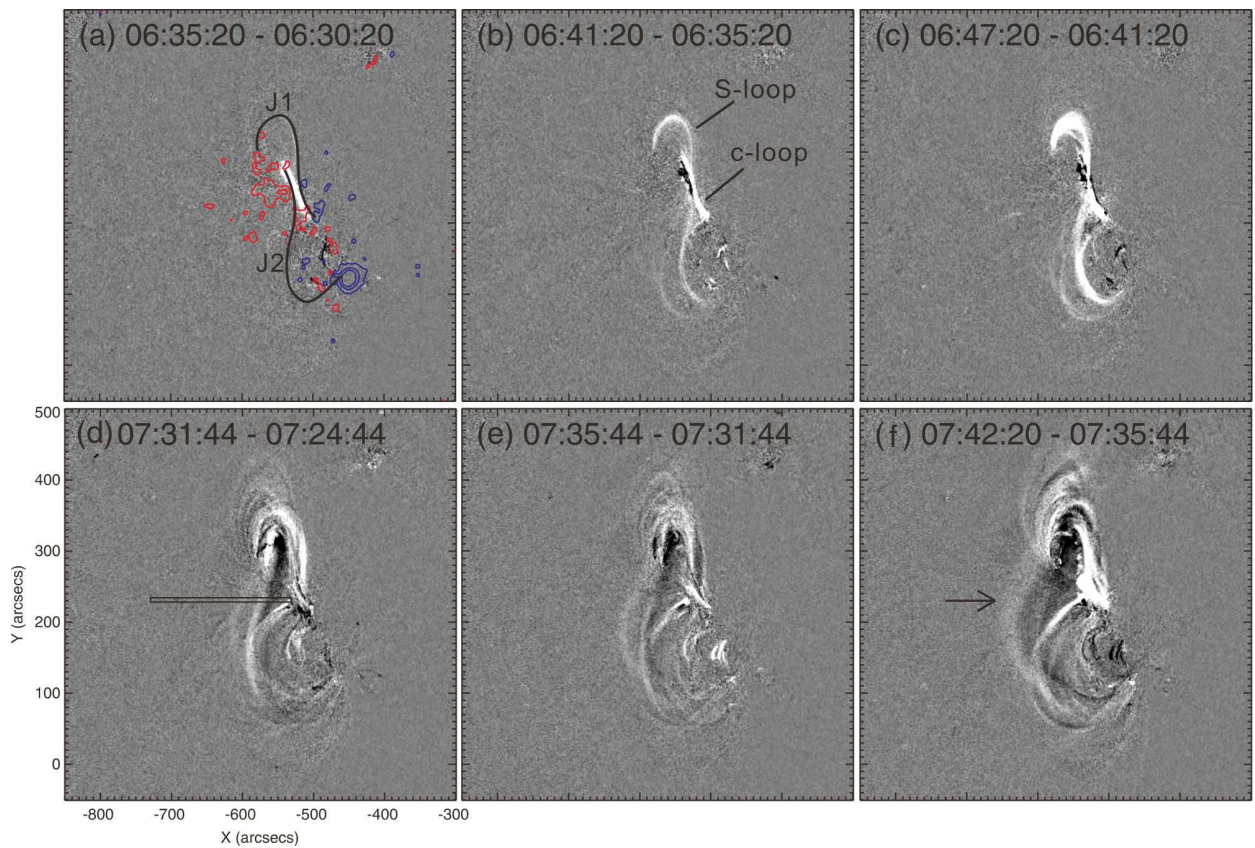}
\caption{AIA 94 \AA\ difference images, displaying the formation of a S-shaped loop via
tether-cutting from two J-shaped loops and its subsequent transformation into a blowing-out
bubble marked by an arrow in Panel (f). Panel (d) shows the slit through which the space-time
diagram in \fig{ltc0801}(c) is obtained. \label{aia0801}}
\end{figure}

The eruption in the sigmoidal region NOAA AR 11093 on 2010 August 1 conformed to the classical
``sigmoid-to-arcade'' transformation \citep[e.g.][]{moore01}, i.e., prior to the eruption, the
sigmoidal structure was consisted of two opposite bundles of J-shaped loops, and after the
eruption, it appeared as a conventional post-flare arcade. The evolution in between the two
states was revealed in detail for the first time by AIA observations \citep{liu10b}. In the AIA
94~\AA\ difference images (\fig{aia0801}), one can see that an S-shaped loop started to glow at
about 06:40 UT, about 1~hr before the flare onset. As its glowing was preceded by a heating
episode in the core region (\fig{aia0801}(a)), the topological reconfiguration resulting in the
formation of the continuous S-shaped loop was very likely due to the tether-cutting reconnection
\citep{moore01}. The S-shaped loop remained in quasi-equilibrium in the lower corona for about 50
minutes, with the central dipped portion rising quasi-statically. During this interval, there was
a weak enhancement in \sat{GOES} soft X-rays (SXRs), whose source, however, was located at the
southeast limb according to \sat{rhessi} observations \citep[see Figure 3 in][]{liu10b}. At about
07:30 UT, about 10 minutes prior to the onset of the C3.2 flare, the speed increased to tens of
kilometers per second, as the S-shaped loop sped up its transformation into an arch-shaped loop,
which eventually led to a CME.

\begin{figure}\epsscale{0.9}
\plotone{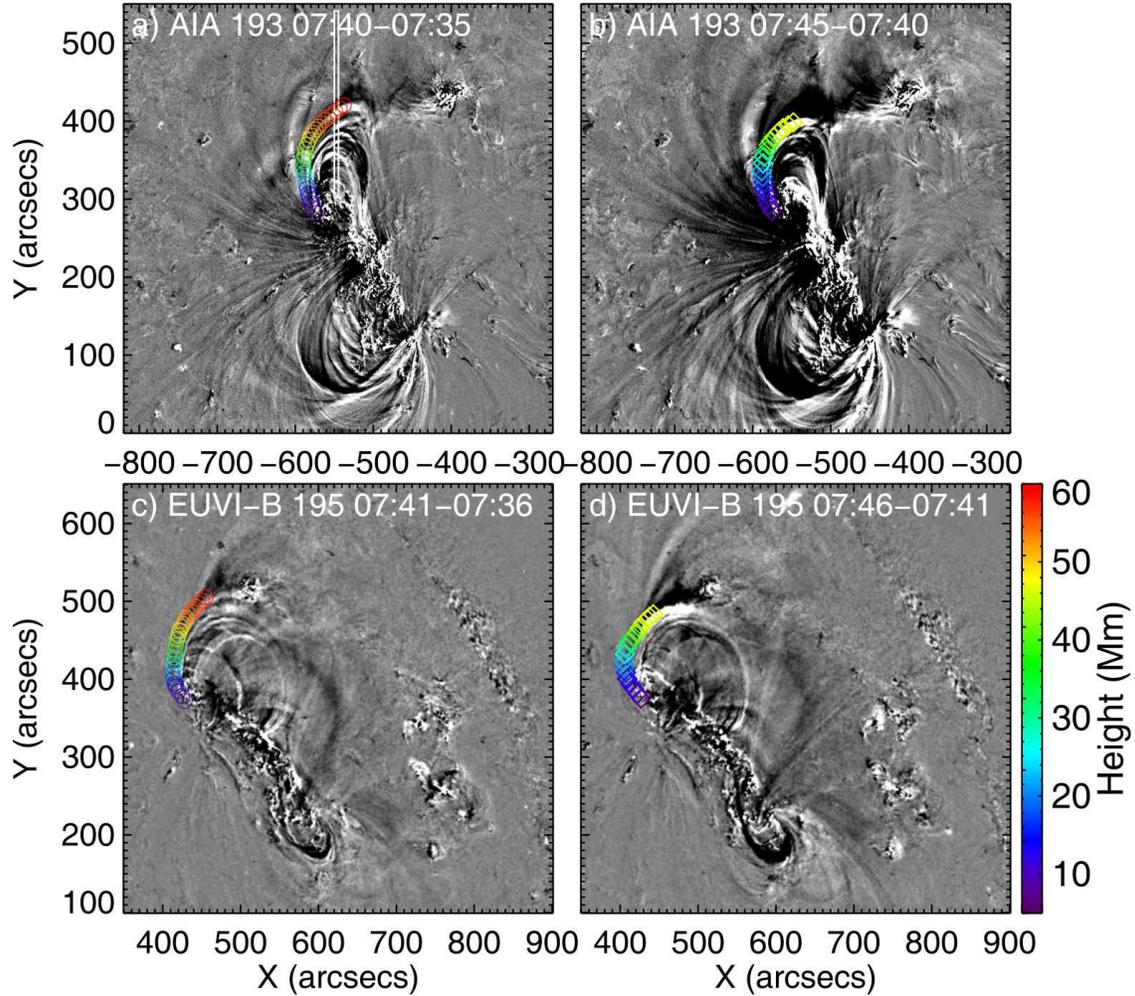}
\caption{Stereoscopic reconstruction of the contracting loop overlying the northern elbow of the
sigmoid. The height information of the loop, which is color-coded, is obtained by pairing AIA 193
\AA\ and EUVI-B 195 {\AA} images. Panel (a) shows the slit through which the space-time diagram
in \fig{ltc0801}(b) is obtained. The expanding bubble is also visible in both view points,
associated with coronal dimming in AIA 193 {\AA}. \label{stereo}}
\end{figure}

\begin{figure}\epsscale{0.8}
\plotone{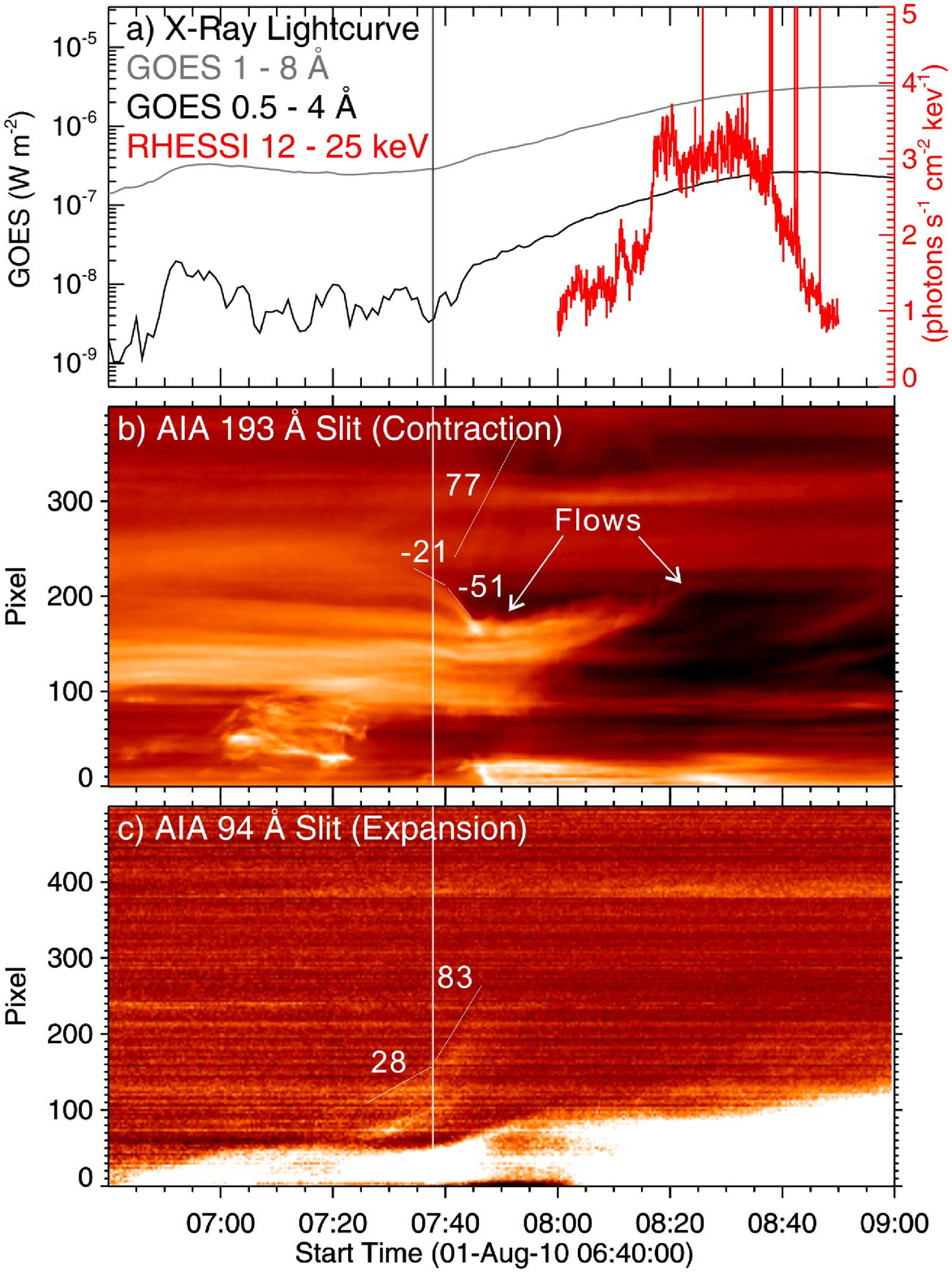}
\caption{Temporal evolution of the contracting loop and the expanding bubble seen through the
slits, in relation to the X-ray emission. Numbers indicate speeds of various features in
km~s${}^{-1}$. The vertical line marks the transition of the exploding bubble from a slow- to a
fast-rise phase. \label{ltc0801}}
\end{figure}

During the eruption, a group of coronal loops overlying the northern elbow of the sigmoid was
observed to contract in cold AIA channels such as 171 and 193 \AA. The contraction was also
visible in EUV images taken by the Extreme-UltraViolet Imager (EUVI; \citealt{wuelser04}) onboard
the ``Behind'' satellite of the Solar Terrestrial Relations Observatory (\sat{stereo-b}). The
viewing angle was separated by about $70^\circ$ between \sat{sdo} and \sat{stereo-b}. By pairing
EUVI with AIA images, we are able to derive the three-dimensional location of the loop undergoing
contraction via a triangulation technique called \emph{tie point} \citep{inhester06}, which is
implemented in an SSW routine, \texttt{SCC\_MEASURE}, by W.~Thompson. From the difference images
(\fig{stereo}) one can see both the contracting loop, whose height is color-coded, and the
expanding bubble, which is associated with coronal dimming in AIA 193 \AA. With stereoscopic
views, it becomes clear that the contraction is not simply a projection effect due to the loops
being pushed aside by the expanding bubble.

We place slits across both the contracting loops (\fig{stereo}(a)) and the expanding bubble
(\fig{aia0801}(d)). By stacking the resultant image cut over time, we obtain space-time diagrams
for a series of AIA 193 and 94 \AA\ images at 12-s cadence (\fig{ltc0801}(b) and (c)). Note that
to increase the signal-to-noise ratio, we integrate over the width of the slit (10 pixels), and
that to reveal the diffuse, expanding bubble, we carry out base difference to make the 94 \AA\
space-time diagram, whereas original 193~\AA\ images are used for the contracting loops which are
more clearly defined in EUV. From \fig{ltc0801}(c), one can see that the bubble initially rose
slowly by \aspeed{30}, and then transitioned into a phase of fast rise by \aspeed{80} at about
07:38 UT. The height-time profile is piecewise-linear fitted although the transition is smooth
and there seems to be a continuous acceleration. The transition time is approximately coincident
with the flare onset in terms of the \sat{goes} 1--8 \AA\ flux. A very diffuse erupting feature
can also be marginally seen in the 193 \AA\ space-time diagram, whose speed is similar as the
bubble in 94 \AA. The contraction of the overlying loops slightly lagged behind the rising of the
bubble, and there was a similar transition from slow to fast contraction, slightly lagging behind
the speed transition of the bubble by less than 3 min. It is worth noting that the apparently
upward-moving feature in the wake of contraction was due to flows along the northern elbow of the
sigmoid, not to the recovery of the contracting loops.

\subsection{2010 September 3 Event}
\begin{figure}\epsscale{1}
\plotone{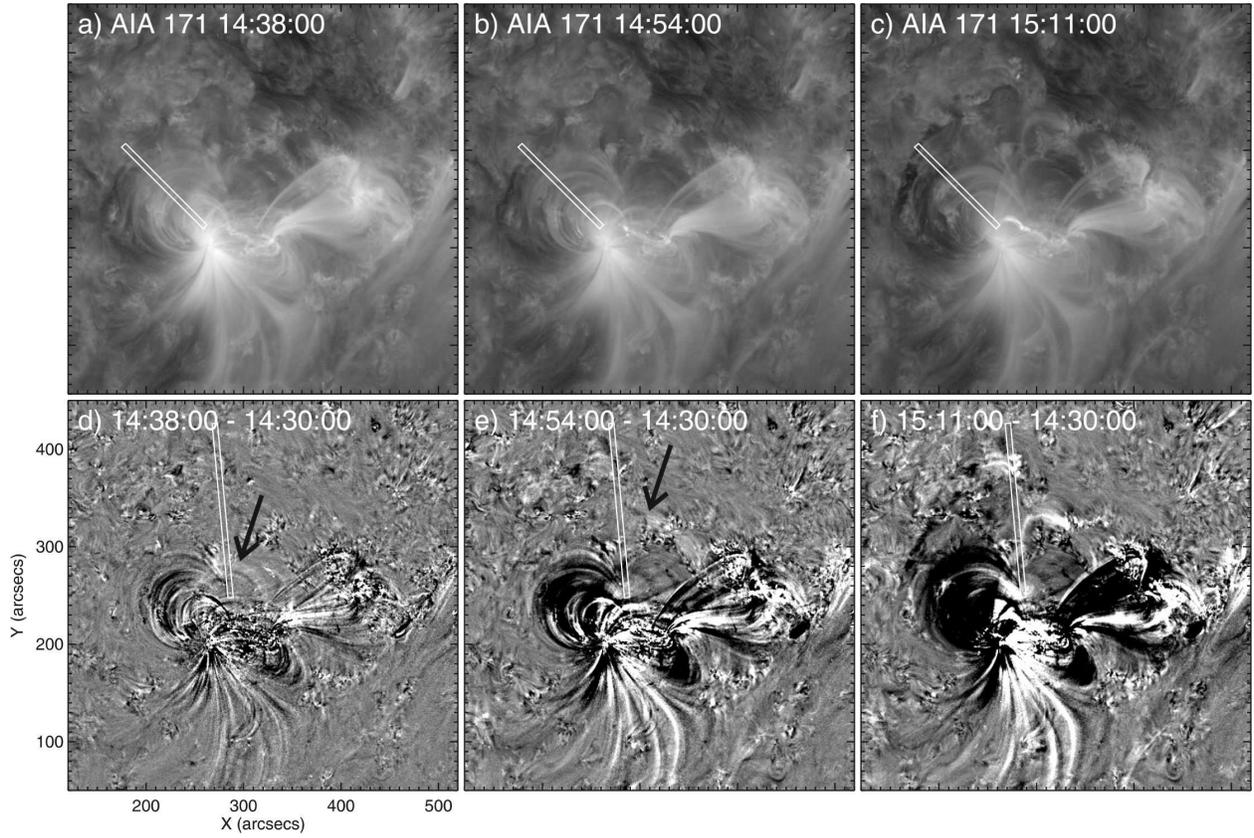} \caption{AIA observation of the 2010 September 3 B-flare. Top panels
show original 171~\AA\ images and bottom panels show the corresponding difference images. The
expanding bubble is indicated by arrows. \label{aia0903}}
\end{figure}

\begin{figure}\epsscale{0.8}
\plotone{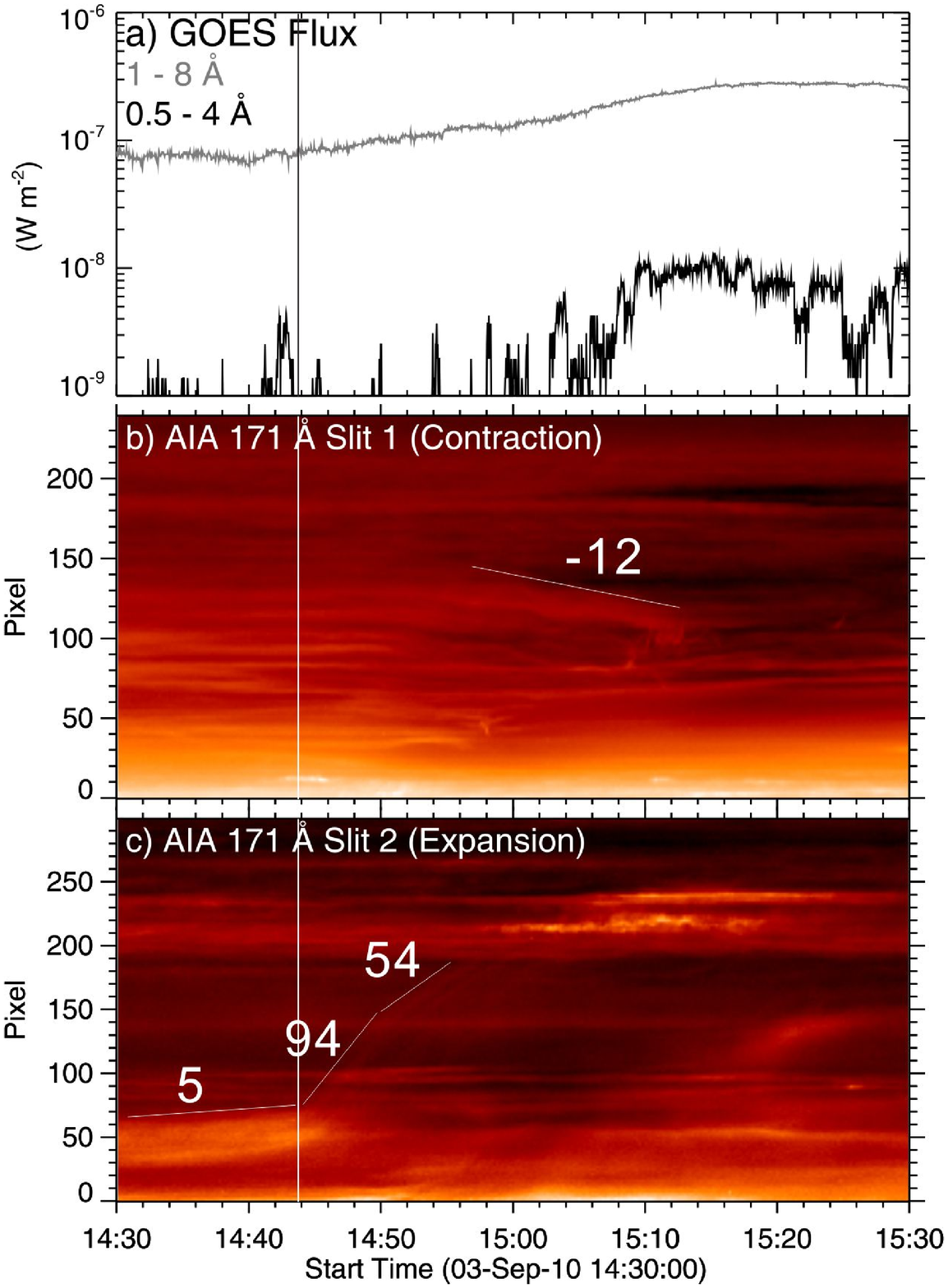} \caption{Temporal evolution of the contracting loop and the expanding
bubble in relation to the X-ray emission. The space-time diagrams are obtained by stacking image
slices cut by the slits shown in \fig{aia0903}. The vertical line marks the transition of the
exploding bubble from a slow- to a fast-rise phase.  \label{ltc0903}}
\end{figure}

In the 2010 September 3 event, both the contracting loops and expanding bubble were visible in the
171 \AA\ channel. But the diffuse bubble can only be seen in difference images (bottom panels in
\fig{aia0903}, marked by arrows). The contracting loops were located to the east of the bubble,
overlying the eastern elbow of the sigmoid (top panels in \fig{aia0903}). Similar to the 2010
August 1 event, in the wake of the bubble erupting, obvious coronal dimming can be seen in the
cold AIA channels such as 171 and 193 \AA. The dynamics of the bubble can also be characterized
by a slow-rise followed by a fast-rise phase, the transition of which coincided with the gradual
increase of the \sat{goes} 1--8~\AA\ flux (\fig{ltc0903}). The bubble shows signature of
deceleration after 14:48 UT. The loop contraction lagged behind the transition time at about
14:44 UT by about 10 minutes.

\subsection{2011 February 13 Event}
\begin{figure}\epsscale{1}
\plotone{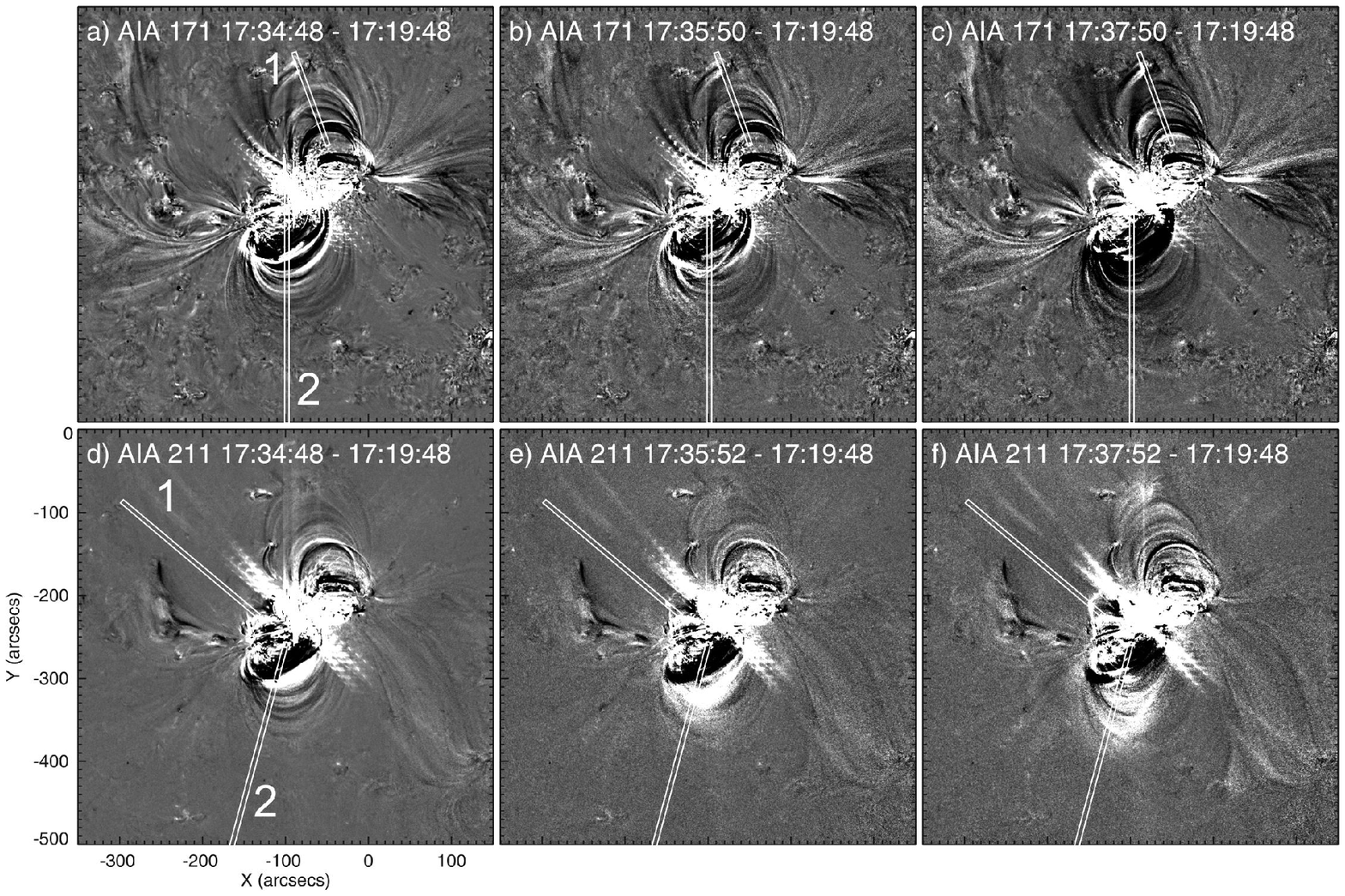} \caption{AIA observation of the 2011 February 13 M-flare. Top panels
show 171~\AA\ difference images and bottom panels 211~\AA\ difference images.  An animation of
211~\AA\ images as well as corresponding difference images is available in the electronic version of the Journal. \label{aia0213}}
\end{figure}

The 2011 February 13 M6.6 flare was associated with irreversible changes of the photospheric
magnetic field \citep{liuc12}. Using high-resolution and high-precision \textit{Hinode} vector
magnetograms and line-of-sight HMI magnetograms, \citet{liuc12} found that the field change
mainly took place in a compact region lying in the center of the sigmoid, where the strength of
the horizontal field increased significantly across the time duration of the flare. Moreover, the
near-surface field became more stressed and inclined toward the surface while the coronal field
became more potential. An intriguing observation is that the current system derived from the
extrapolated coronal field above the region with enhanced horizontal field underwent an apparent
downward collapse in the wake of the sigmoid eruption. \citet{liuc12} concluded that these
results are a superimposed effect of both the tether-cutting reconnection producing the flare and
the magnetic implosion resulting from the energy release.

\begin{figure}\epsscale{0.55}
\plotone{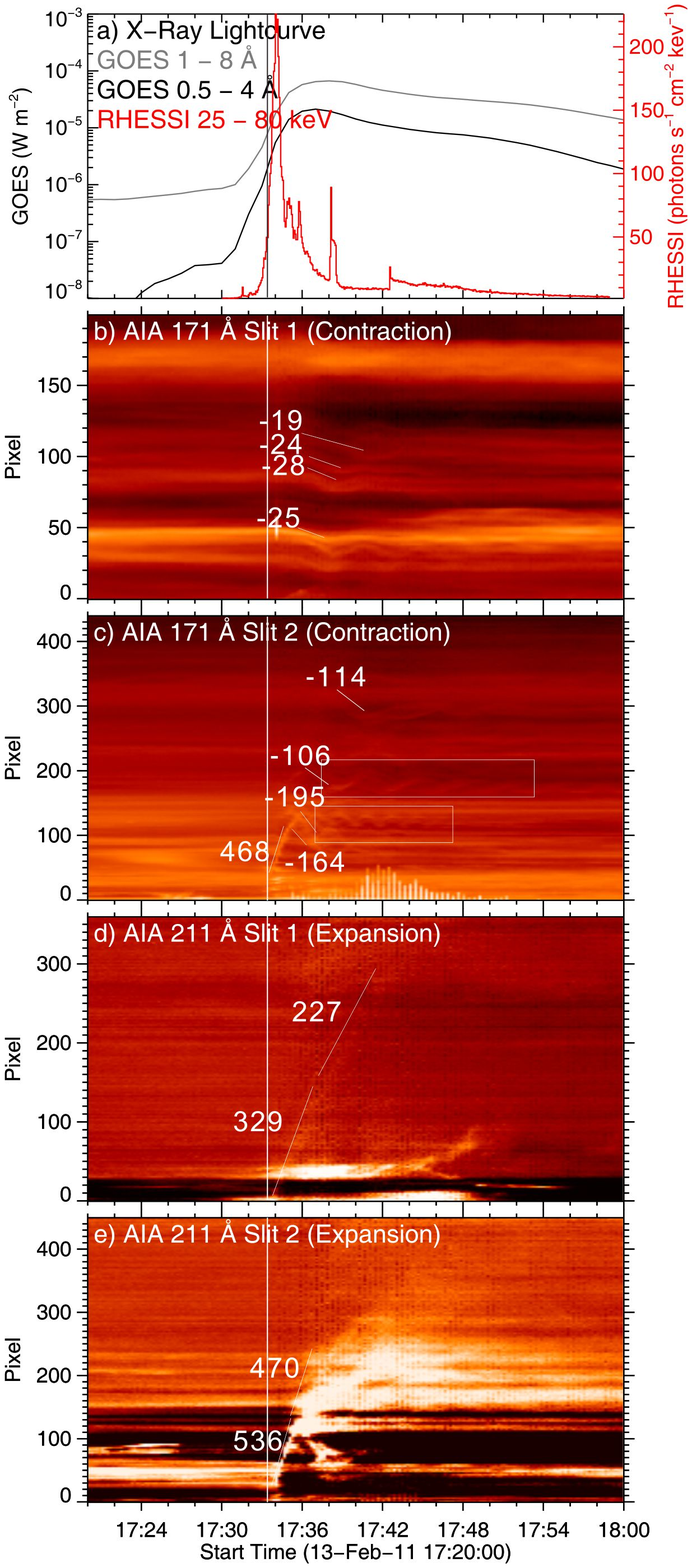} \caption{Temporal evolution of the contracting loop and the expanding
bubble in relation to the X-ray emission. The space-time diagrams are obtained by stacking image
slices cut by the slits as shown in \fig{aia0213}. The vertical line marks the beginning of the
explosion.\label{ltc0213}}
\end{figure}

\begin{figure}\epsscale{1}
\plotone{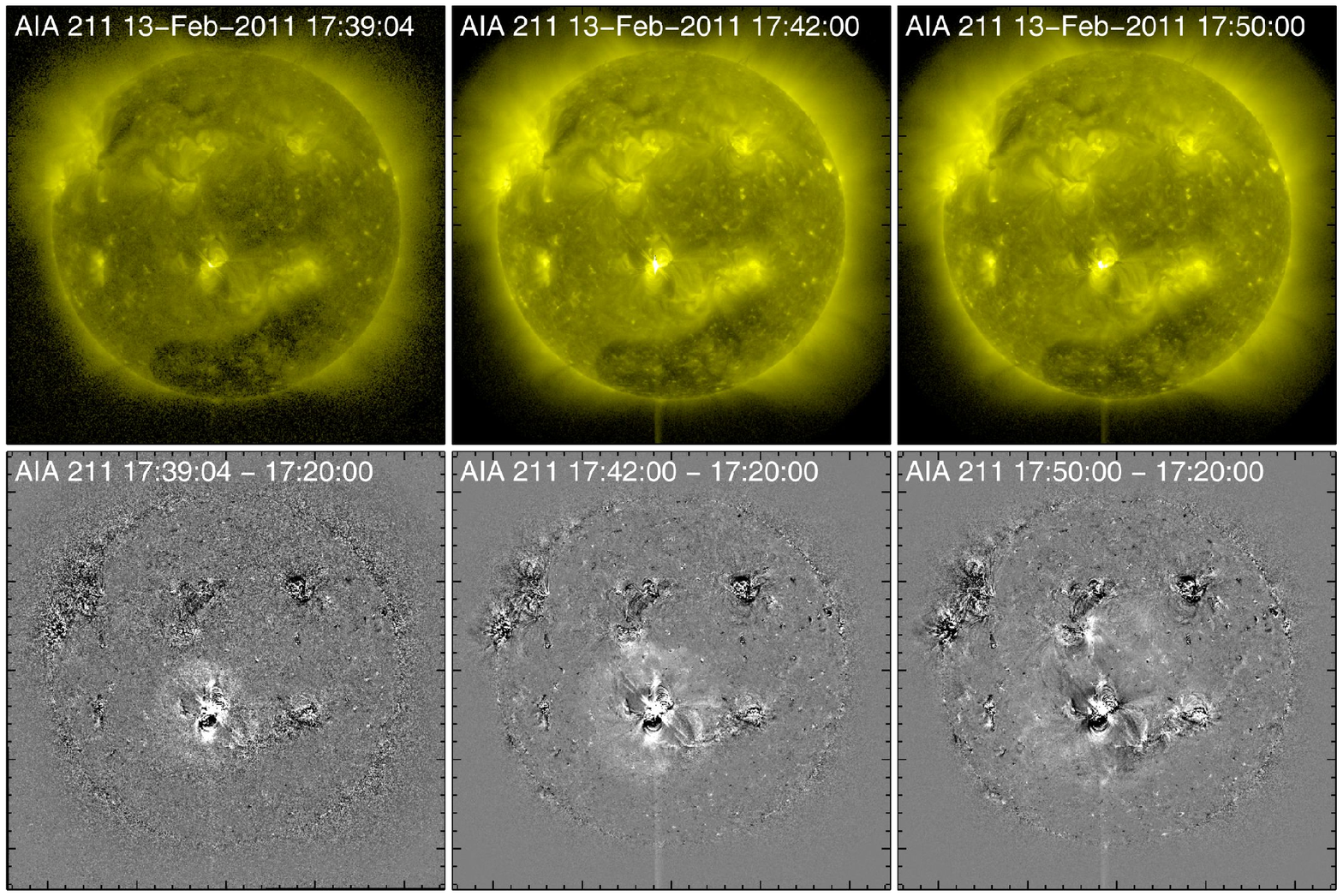}
\caption{Snapshots of the full-disk AIA 211 \AA\ images (top panels) and the corresponding
difference images (bottom panels). In the difference images, a diffuse front can be seen to
propagate outward from the active region of interest. \label{wave}}
\end{figure}

Coronal EUV observations agree with the above conclusion drawn from photospheric field
measurements regarding magnetic implosion. At the onset of the impulsive phase, two arch-shaped
loops originating from the center of the sigmoid were observed to expand outward in 211~\AA\ in
different directions (bottom panels of \fig{aia0213}) but at similar projected speeds
(\fig{ltc0213}(d) and (e)), while coronal loops overlying both elbows of the sigmoid were
observed to contract (top panels of \fig{aia0213}), with the loops overlying the eastern elbow
contracting much faster (\fig{ltc0213}(b) and (c)). For this relatively energetic event, the
eruption only preceded the contraction by tens of seconds, and the contracting speed reaches as
fast as \speed{200}. In the wake of the contraction, loops overlying the eastern elbow underwent
oscillation for several cycles (marked by rectangles in \fig{ltc0213}), similar to the events
studied by \citet{lw10}, \citet{gosain12}, and \citet{kp12}. Beyond the expanding loops, one can also see in the animation of AIA 211
\AA\ difference images (accompanying \fig{aia0213}) a diffuse oval front with enhanced intensity
propagating outward, well separated from the expanding loops. This oval structure has been
identified in MHD simulations as a shell of return currents surrounding the flux rope
\citep{aulanier10, schrijver11}. From \fig{wave} one can see that the front was propagating
anisotropically, apparently restrained by nearby active regions and the coronal hole in the
southern polar region.

\subsection{2011 February 15 Event}
The 2011 February 15 X2.2 flare in AR 11158 is the first X-class flare of the current solar
cycle, hence it raises a lot of interests and has been intensively studied. \citet{sasha11}
reported that the flare produced a powerful ``sunquake'' event with its impact on the
photosphere. \citet{wangs12} reported a rapid, irreversible change of the photospheric magnetic
field associated with the flare. \citet{bvd12} studied the shear flows along the
polarity-inversion line as well as the white-light flare emission. \citet{schrijver11} investigated the
coronal transients associated with the flare. In particular, \citet{schrijver11} observed
``expanding loops from a flux-rope-like structure over the shearing polarity-inversion line
between the central $\delta$-spot groups of AR 11158, developing a propagating coronal front
(`EIT wave'), and eventually forming the coronal mass ejection moving into the inner
heliosphere''. Here the expanding loops are identified as the erupting component of the
eruption.

\begin{figure}\epsscale{1}
\plotone{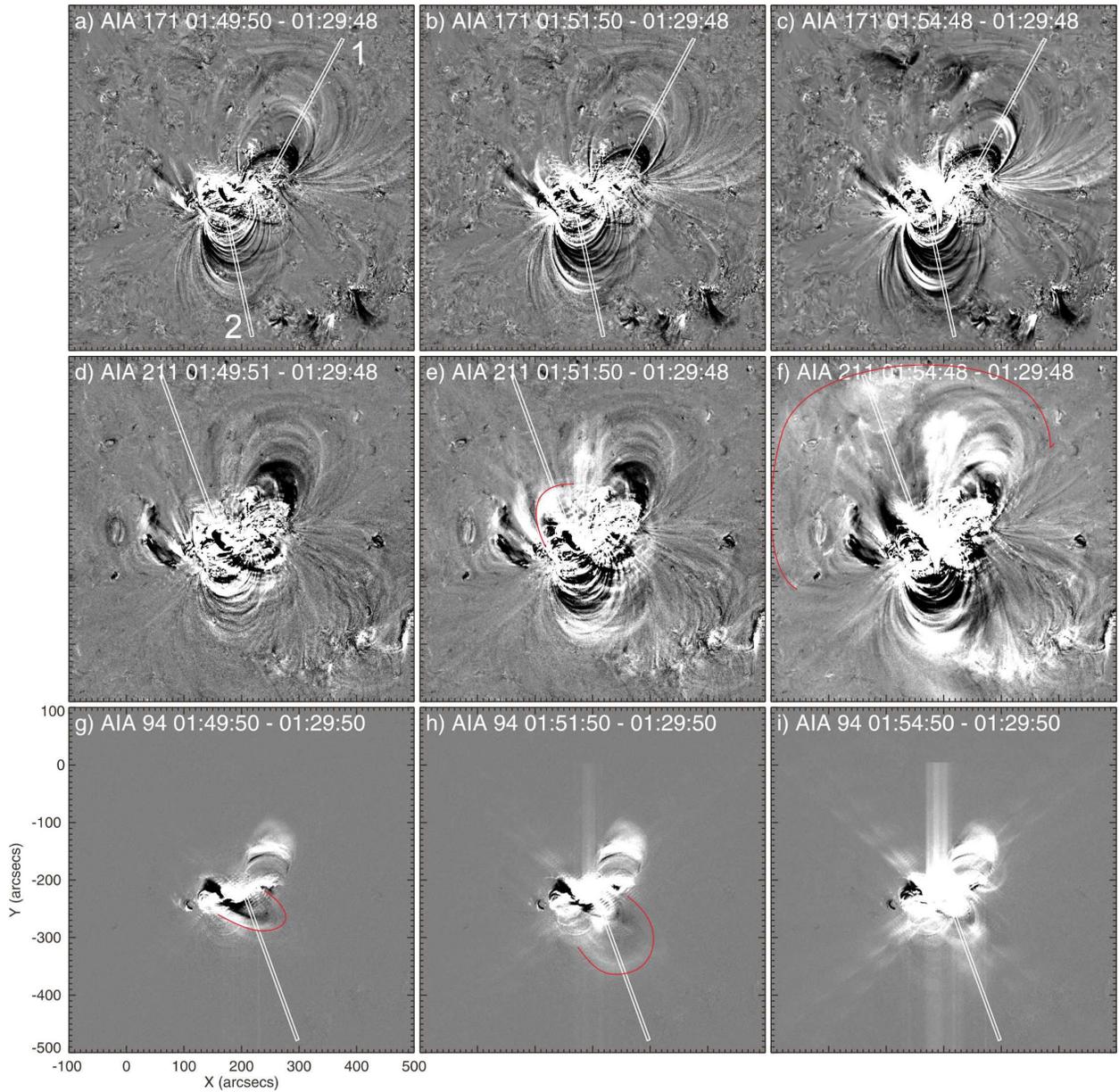} \caption{AIA observation of the 2011 February 15 X-flare. From top,
middle to bottom panels, we show 171, 211, and 94 difference images, respectively. The expanding
bubble is highlighted by red arcs. \label{aia0215}}
\end{figure}

\begin{figure}\epsscale{0.55}
\plotone{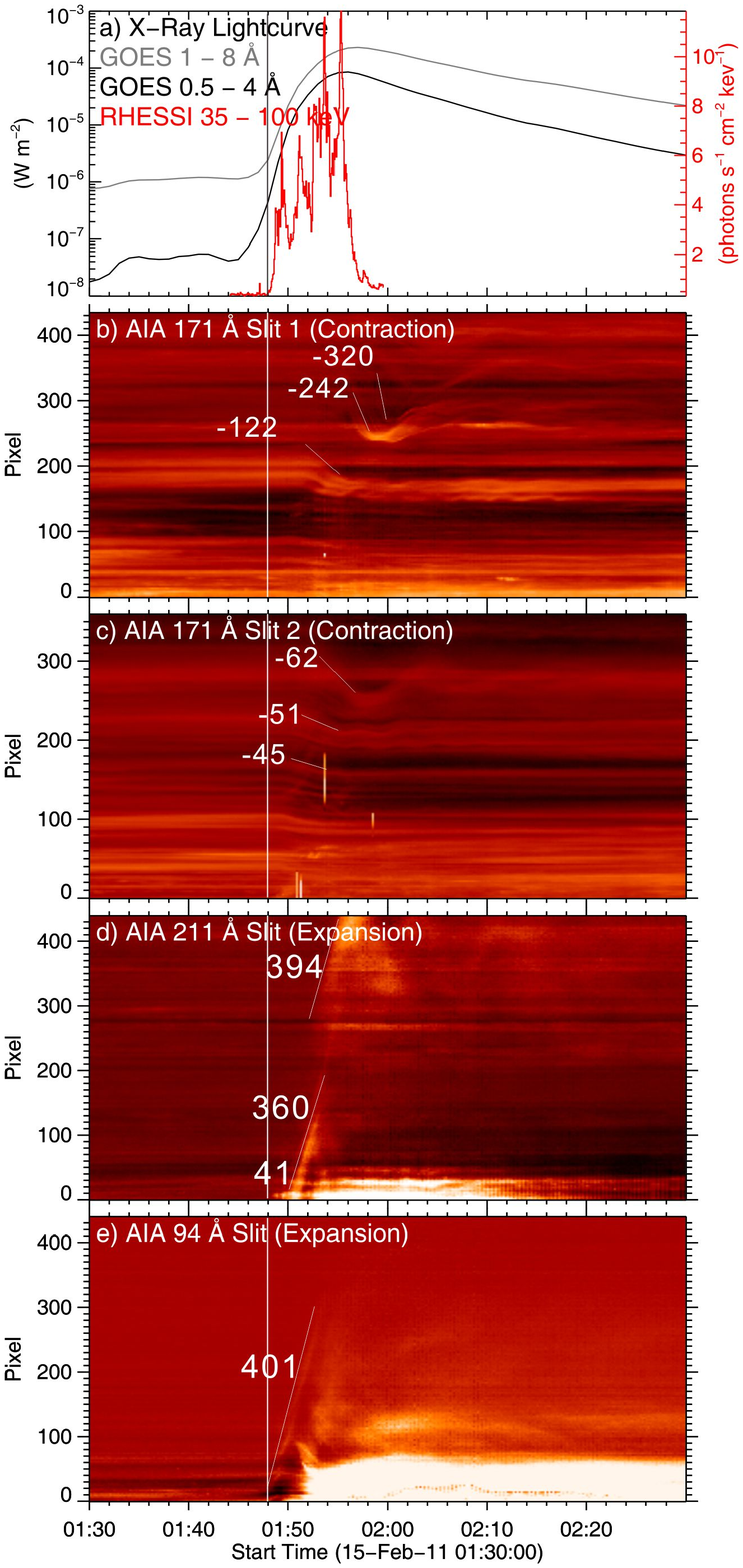} \caption{Temporal evolution of the contracting loop and the expanding
bubble in relation to the X-ray emission. The space-time diagrams are obtained by stacking image
slices cut by the slits as shown in \fig{aia0215}. The vertical line marks the beginning of the
explosion. \label{ltc0215}}
\end{figure}

The active region as seen in the AIA 171 \AA\ channel were dominated by two groups of
potential-like loops overlying the elbows of the forward S-shaped sigmoid as seen in the hot AIA
channels (\fig{sigmoid}). Both groups of potential-like loops were observed to contract during
the X2.2 flare. In each group, loops underwent contraction in a successive fashion with those
located at lower altitudes starting to contract first, due presumably to the limited propagation
speed of the Alf\`{v}en wave \citep[see also][]{lw10,gosain12}, whereas loops at higher altitudes
had a faster contraction speed (\fig{ltc0215}(b) and (c)). These contracting/collapsing features
are also independently noticed by \citet{schrijver11,gosain12,sun12} with different approaches
but a similar interpretation in agreement with \citet{lw09}. Similar to the February 13 event, in
the wake of the contraction, loops overlying both elbows underwent oscillation \citep[see \fig{ltc0215}, 
also see][]{lw10, gosain12, kp12}.


Immediately prior to the loop contraction, a bubble (marked by red arcs in the middle panels of
\fig{aia0215}) can be best seen to originate from the core of the sigmoid and to expand
northeastward in the 211 \AA\ channel and southwestward in the 94 \AA\ channel (bottom panels of
\fig{aia0215}; marked by red arcs). A transition from a slow- to fast-rise phase can still be
marginally seen in the 211~\AA\ channel. But the duration of the slow-rise phase was very short,
lasting for only about 2 minutes. The transition time at about 01:50 UT still preceded the loop
contraction by about 3 minutes. The commencement of the bubble expansion at about 01:48 UT was
concurrent with the onset of the nonthermal HXR emission at 35--100 keV. This expanding bubble
was also closely associated with ``an expanding intensity front propagating away from the flaring
region seen on the disk, and the leading edge of the intensity signature of the CME propagating
outward from the Sun into the heliosphere'' as identified by \citet{schrijver11}. These three
distinct features are suggested as different observational aspects of the eruption of a flux rope
\citep{schrijver11}.

\subsection{2011 June 21 Event}

\begin{figure}\epsscale{1}
\plotone{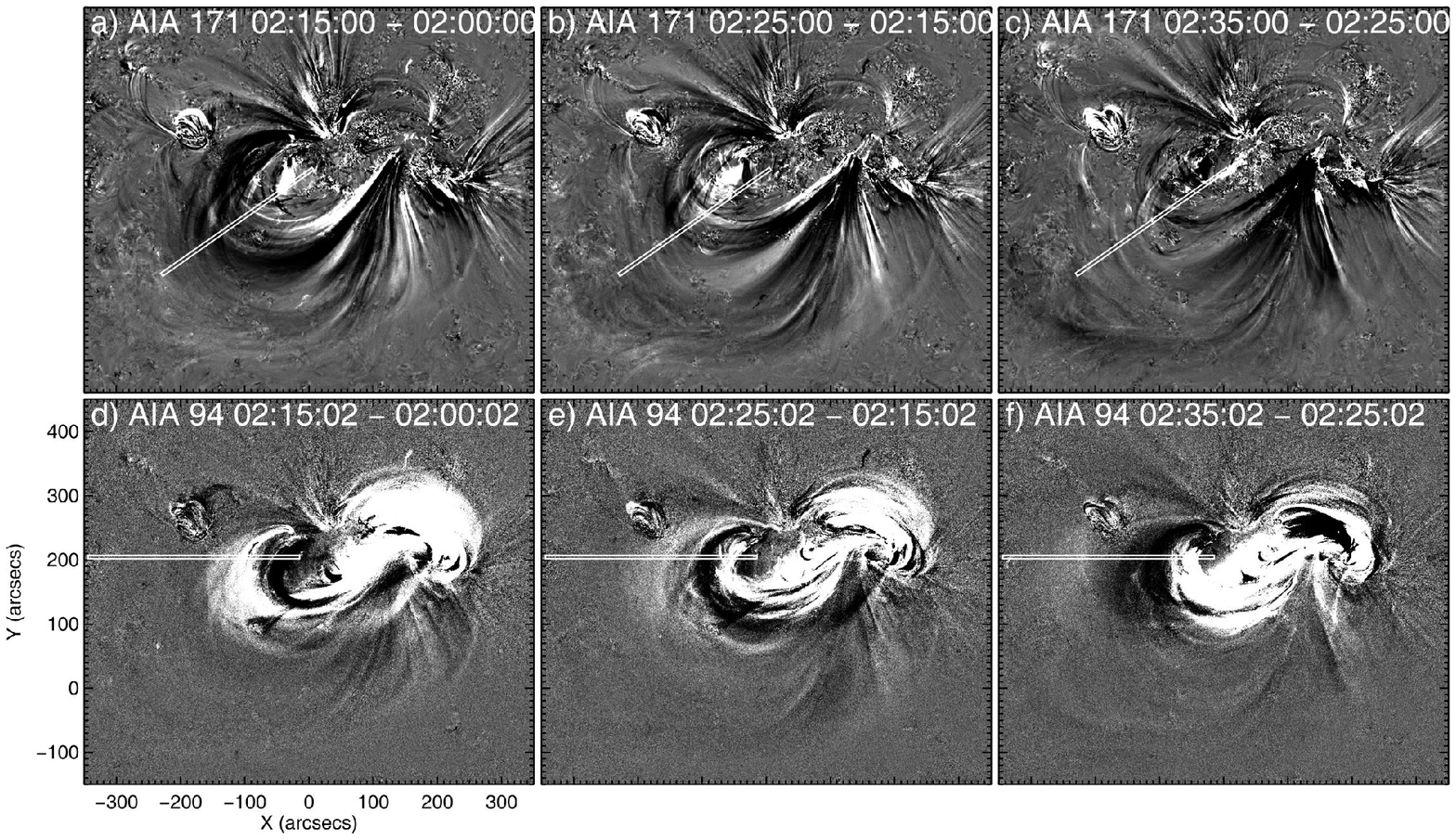} \caption{AIA observation of the 2011 June 21 C-flare. Top panels show
171~\AA\ difference images and bottom panels 94~\AA\ difference images. \label{aia0621}}
\end{figure}

\begin{figure}\epsscale{0.7}
\plotone{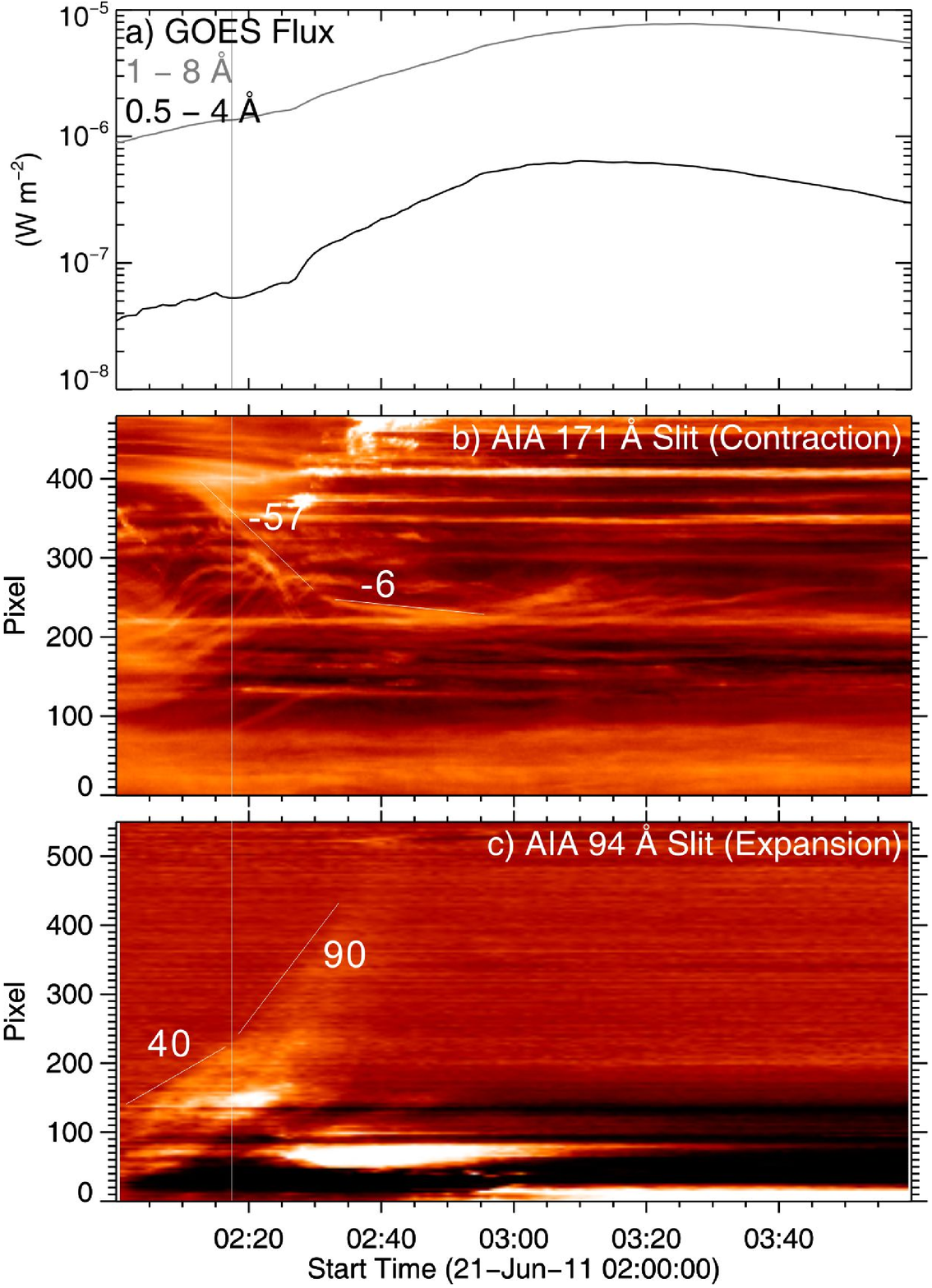}
\caption{Temporal evolution of the contracting loops and the expanding bubble in relation to the
X-ray emission. The space-time diagrams are obtained by stacking image slices cut by the slits as
shown in \fig{aia0621}. The vertical line marks the transition of the exploding bubble from a
slow- to a fast-rise phase. \label{ltc0621}}
\end{figure}

In the 2011 June 21 event, the group of coronal loops overlying the eastern elbow of the sigmoid
was observed to contract in the 171 \AA\ channel (top panels in \fig{aia0621}). At the same time,
a bubble originating from the center of the sigmoid was observed to expand eastward in the 94
\AA\ channel (bottom panels in \fig{aia0621}). Both the contraction and the expansion occurred
prior to the C7.7 flare. The transition time of the bubble from a relatively slow- to a fast-rise
phase was roughly coincident with the onset of the flare (\fig{ltc0621}).

\section{Discussion \& Conclusion}

\begin{figure}\epsscale{0.8}
\plotone{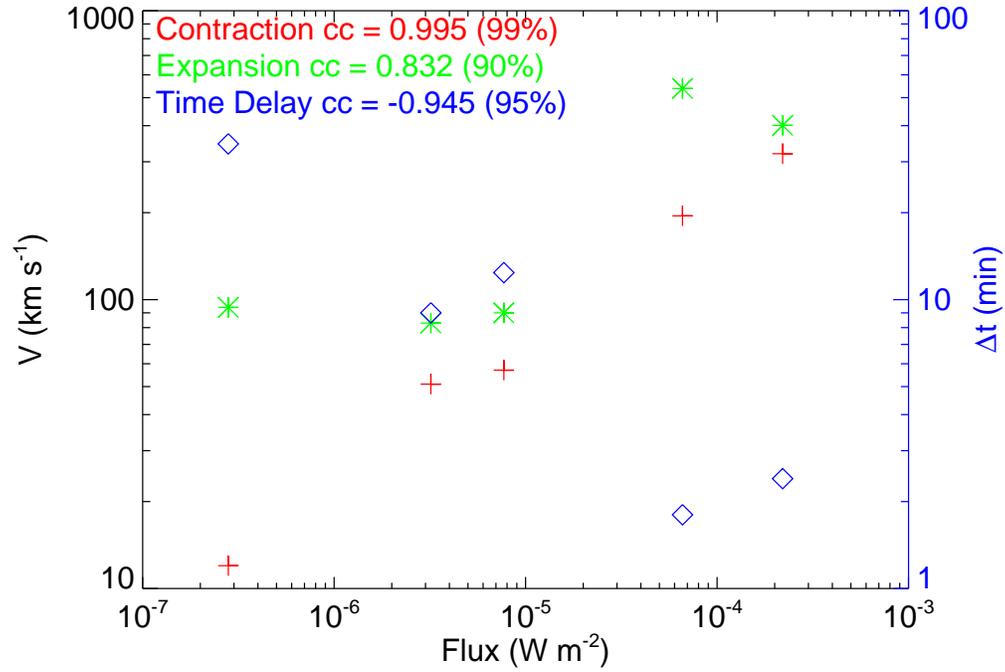} \caption{Correlation of the maximal contraction/expansion speed, $V$, and the
time delay of contraction relative to expansion, $\Delta t$, with the flare magnitude in terms of
the peak \sat{goes} 1--8 \AA\ flux. The confidence level of the linear correlation coefficient,
$cc$, of $\lg(V)$ and $\lg \Delta t$ with $\lg(F)$ is given in the brackets. \label{correl}}
\end{figure}

We have investigated four sigmoidal active regions, in which five eruptions with
signatures of magnetic implosion occurred. The magnitudes of the flares associated with the
eruptions span almost the whole flare ``spectrum'', from the \sat{goes}-class B to X. In all of
the flares studied, there are both a contracting and an erupting component: the former is only
observed in cold EUV channels and the latter is preferentially visible in warm/hot EUV channels.
This is because the contracting component is composed of large-scale, potential-like coronal
loops overlying the elbows of the sigmoid, while the erupting component is associated with newly
reconnected flux tubes originating from the center of the sigmoid \citep[c.f.,][]{liu10b,
aulanier10, schrijver11}. Several important aspects of these observations are discussed as follows.

\begin{itemize}
\item \textbf{Consequence of loop contraction:} the overlying loops undergoing contraction never
regain their pre-flare positions, which implies a new equilibrium with reduced magnetic energy as
the eruption is powered by magnetic energy. One may argue that the apparent contraction of coronal
loops could be a projection effect, i.e., the loop plane tilts due to the flare impulse. But in
that case, one would expect the restoration of the loops once the flare impulse has passed away.
In observation, however, the contracting loops may oscillate about a lower height \citep[e.g.,
\fig{ltc0213}; see also][]{lw10}, but never reach the original heights after the eruption. Thus,
the contraction within the loop plane must make a significant contribution. 

\item \textbf{Correlation between contraction and eruption:} the contraction speed seems to depend
on the intensity/magnitude of the eruption. From \fig{correl}, one can see that despite this very
small sample size, the peak \sat{goes} SXR flux as a proxy of the flare magnitude is linearly
correlated very well with the measured maximal contracting speed in the log-log plot, although not
so well with the maximal erupting speed. Unlike contracting loops which are clearly defined,
however, the measurement of the erupting speed involves larger uncertainties as the front of the
expanding bubble tend to get more and more diluted and eventually overwhelmed by the background
during propagation, thereby leading to underestimation of its speed. One more caveat to keep in
mind is that these speeds are not necessarily measured at the time of the peak SXR flux.


\item \textbf{Timing:} the eruption precedes the contraction in all of the flares studied, which
establishes loop contraction as a consequence of eruption. There is also a trend that the more
energetic the eruption, the smaller the time delay of the loop contraction relative to the onset
of the expansion of the erupting component, which is demonstrated in \fig{correl} as a strong
anti-correlation between the time delay and the peak \sat{goes} SXR flux in the log-log plot.
This time delay is presumably determined by the expansion speed of the erupting component. In
addition, in the relatively weak B- and C-flares, the initiation of the erupting component
precedes the increase in \sat{goes} SXR fluxes; but in the stronger M- and X-flares, it is
concurrent with the increase in nonthermal HXR fluxes. This may lend support to \citet{lin04}, who
concluded that CMEs are better correlated with flares if there is more free energy available to
drive the eruption. On the other hand, since the CME progenitor, i.e., the expanding bubble, forms
before the flare onset as the weak events clearly demonstrate, the CME must be independent of the
conventionally defined flare, or, the flare is only a byproduct of the CME, unless the eruption
mechanism for the weak events is different from that for the energetic ones.

\item \textbf{Asymmetry of contraction:} the two groups of coronal loops overlying the elbows of
the sigmoid often contract asymmetrically, i.e., not only they contract at different speeds but
either group could show little sign of contraction, dependent on the detailed interaction between
the core field and the arcade field, including, presumably, their relative strength and the
spatial distribution of the decay index of the restraining field \citep{kt06, lag09, liuc10}. For
the 2010 August 1 event in particular, \citet{liu10b} concluded that the majority of the flare
loops were formed by reconnection of the stretched legs of the less sheared loops overlying the
southern elbow and the center of the sigmoid, based on the reconnection rate inferred from the
H$\alpha$ ribbon motion. The eruption therefore left most loops overlying the northern elbow
unopened. This explains why only these loops underwent obvious contraction. The
intensity/magnitude of the eruption could be another relevant factor as among the events studied only
those greater than M-class show contraction of loops overlying both elbows of the sigmoid. 

\item \textbf{Implication for eruption mechanism:} as the contracting component is distinct from
the erupting component, we conclude that these eruptions conform to the ``rupture model'' in
which the arcade field is partially opened \citep[][\fig{model}(b)]{sturrock01}. We can further
exclude the breakout model because the coronal loops undergoing contraction are arched over,
rather than located to the side of, the sheared core field. The loop contraction in the latter
occasion results from reconnection at the magnetic null above the central lobe in the breakout
model.
\end{itemize} 

In conclusion, these observations substantiate the loop contraction as an integrated process in
eruptions of sigmoidal active regions in which the restraining arcade field is only partially
opened, consistent with theoretical expectations. The consequence of loop contraction is a new
equilibrium of the coronal field with reduced magnetic energy, and the process itself is a result
of the flare energy release, as evidenced by the strong correlation of the maximal
contracting speed, and strong anti-correlation of the time delay of contraction relative to
expansion, with the peak SXR flux.

\acknowledgments The authors are grateful to the \sat{sdo}, \sat{stereo} and \sat{rhessi} teams
for the free access to the data and the development of the data analysis software. R.L.
acknowledges the Thousand Young Talents Program of China. R.L. and Y.W. were supported by grants
from NSFC 41131065 and 41121003, 973 key project 2011CB811403, CAS Key Research Program
KZZD-EW-01-4, and the fundamental research funds  for the central universities WK2080000031.
R.L., C.L. and H.W. were supported by NSF grants ATM-0849453 and ATM-0819662. The contribution of
T.T. was supported by CISM (an NSF Science and Technology Center).


\end{document}